\documentclass[twocolumn,10pt]{asme2ej}

\usepackage{lmodern}
\usepackage{lineno}
\usepackage[percent]{overpic}
\usepackage{amsmath}
\usepackage{nicefrac}
\usepackage[aboveskip=3pt]{subcaption}
\usepackage{siunitx}
\usepackage{xcolor,varwidth}
\usepackage{multirow}
\usepackage{graphicx} %% for loading jpg figures
\usepackage{tabularx}

\title{Interaction of void spacing and material size effect on inter-void flow localisation}

\author{Ingrid Holte
\affiliation{
	Department of Mechanical Engineering\\
	Technical University of Denmark, Kgs. Lyngby, Denmark\\
    E-mail: inghol@mek.dtu.dk
}}

\author{Ankit Srivastava
\affiliation{
Department of Materials Science and Engineering\\
Texas A\&M University, College Station Texas, USA\\
E-mail: ankit.sri@tamu.edu
}}

\author{Emilio Mart$\textbf{\'\i}$nez-Pa$\tilde{\textbf{n}}$eda
\affiliation{
Department of Civil and Environmental Engineering\\ 
Imperial College, London, UK\\
E-mail: e.martinez-paneda@imperial.ac.uk 
}}

\author{Christian F. Niordson$^1$,
Kim L. Nielsen$^2$
\affiliation{ 
Department of Mechanical Engineering\\
Technical University of Denmark, Kgs. Lyngby, Denmark\\
$^1$E-mail: cn@mek.dtu.dk  \\
$^2$E-mail: kin@mek.dtu.dk
}}

\begin{document}

\maketitle

%%%%%%%%%%%%%%%%%%%%%%%%%%%%%%%%%%%%%%%%%%%%%%%%%%%%%%%%%%%%%%%%%%%%%%
\begin{abstract}
{\it The ductile fracture process in porous metals due to growth and coalescence of micron scale voids is not only affected by the imposed stress state but also by the distribution of the voids and the material size effect. The objective of this work is to understand the interaction of the inter-void spacing (or ligaments) and the resultant gradient induced material size effect on void coalescence for a range of imposed stress states. To this end, three dimensional finite element calculations of unit cell models with a discrete void embedded in a strain gradient enhanced material matrix are performed. The calculations are carried out for a range of initial inter-void ligament sizes and imposed stress states characterised by fixed values of the stress triaxiality and the Lode parameter. Our results show that in the absence of strain gradient effects on the material response, decreasing the inter-void ligament size results in an increase in the propensity for void coalescence. However, in a strain gradient enhanced material matrix, the strain gradients harden the material in the inter-void ligament and decrease the effect of inter-void ligament size on the propensity for void coalescence.}
\end{abstract}
%%%%%%%%%%%%%%%%%%%%%%%%%%%%%%%%%%%%%%%%%%%%%%%%%%%%%%%%%%%%%%%%%%%%%%

\section{Introduction}

In porous metals, void coalescence often drives the onset of the macroscopic flow localisation that marks the end of uniform deformation and acts as a precursor to failure, as well as the initiation and propagation of ductile cracks \cite{tekouglu2015localization, guo2018void, liu2019micromechanism}. Previous studies suggest that for conventional plasticity theory, where no material length scale enters the constitutive law (absence of stress/strain gradient induced size effect), a decrease in the inter-void spacing promotes void coalescence \cite{pardoen2000extended, srivastava2013void} and results in the collapse of the yield surface \cite{torki2015void, torki2017theoretical}. While for a fixed inter-void spacing, it is well established that the imposed stress state has a pronounced effect on the onset of void coalescence in conventional plasticity theory. For example, it has been shown that an increase in the imposed stress triaxiality (a ratio of the first to second stress invariant) promotes void growth and early onset of void coalescence \cite{budiansky1982void, koplik1988void, needleman1992void, benzerga2010ductile}. Void coalescence is simply the event where the plastic flow localises within the inter-void ligaments and successively links the neighboring voids \cite{koplik1988void}. The plastic flow localisation within the inter-void ligament, however, will induce plastic strain gradients that in turn may affect the strengthening and hardening of the material. This raises a fundamental question: how does the interaction of inter-void spacing (or ligament size) and the gradient induced material size effect, affect the localisation of plastic flow causing void coalescence for a given stress state?        

The gradient induced size effect resulting in strengthening and hardening in metals has been confirmed in many material tests involving non-uniform deformation including indentation \cite{stelmashenko1993,ma1995size}, torsion \cite{fleck1994}, and bending \cite{stolken1998}. The size dependent material response on the micron scale in metal plasticity implies that the growth of micron sized voids also exhibits significant size effects \cite{tvergaardniordson_2007, niordson_2008}. In general, it has been shown that the gradient induced size effect leads to slower growth rates for smaller voids \cite{tvergaard2004nonlocal, li2006rve, monchiet2013gurson, nielsen_2016, holte2019}. An accurate representation of void coalescence due to plastic flow localisation within micron sized inter-void ligaments, therefore, also requires material models that represent stresses over the relevant length scales. Phenomenological theories describing the strengthening and hardening due to plastic strain gradients express the plastic work in terms of both plastic strain and plastic strain gradient, thereby introducing a length scale into the material model. Herein, the strain gradient plasticity theory proposed by Gudmundson \cite{gudmundson2004} is used, which includes both dissipative (non-recoverable) and energetic (recoverable) gradient contributions within a small strain formulation based on visco-plasticity. The mathematical formulation and associated variational structure originate from Fleck and Willis \cite{fleck2009}, and the material model is implemented into the commercial finite element software ABAQUS using a user element (UEL) subroutine \cite{martinez2019}.

The objective of this work is to understand the interaction of the inter-void spacing (or ligament size) and the resultant gradient induced material size effect on void coalescence for a range of imposed stress states. To achieve this, three dimensional finite element unit cell calculations for a periodic array of initially spherical voids embedded in a strain gradient enhanced material matrix are carried out. Several unit cell geometries have been analyzed to investigate the effect of inter-void ligament size under multiple loading conditions. The imposed stress states are characterised by fixed values of the stress triaxiality and the Lode parameter (a measure of the third stress invariant). The value of the Lode parameter is shown to affect the evolution of voids in computations involving conventional plasticity theory \cite{zhang2001numerical, kim2004modeling, gao2006modeling, barsoum2007rupturemicro, srivastava2013void} and in experiments \cite{bao2004fracture, barsoum2007rupture, srivastava2012effect} only at relatively low stress triaxiality levels. However, it is likely that in an anisotropic material matrix \cite{srivastava2015effect} with anisotropy introduced by the void distribution \cite{srivastava2013void}, as for the present investigation, the effect of the Lode parameter can be important even at high stress triaxialities.

Our results show that for a conventional material matrix, increasing the inter-void ligament size results in an increase in the critical stress to void coalescence, up to a threshold value of inter-void ligament size. The sensitivity of the critical stress to the inter-void ligament size is found to increase with increasing stress triaxiality. The quantitative effect of the Lode parameter is found to be small for the stress triaxiality values varying from $1$ to $3$. However, for inter-void ligament sizes below the threshold value, the critical stress is smallest for a Lode parameter value of $-1$, whereas above the threshold value the critical stress is smallest for a Lode parameter value of $0$. For a void in a strain gradient enhanced material matrix, the value of the critical stress for void coalescence increases with increasing length parameter i.e. increasing gradient effect. This effect of the length parameter on the critical stress magnitude is found to increase with increasing imposed stress triaxiality and decreasing inter-void ligament size. This is because at higher stress triaxiality values and for smaller inter-void ligament sizes, there is an increase in the propensity for plastic flow localisation that introduces strong plastic strain gradients and in turn hardens the ligament. This mechanism leads to a decrease in the dependence of critical stress on the inter-void ligament size with increasing length parameter. The gradient induced strengthening also tends to homogenize the deformation in the unit cell thus decreasing the effect of the Lode parameter. 

The structure of the manuscript is as follows. Section~\ref{S:problemformulation} frames the study and presents the numerical method. The unit cell geometries considered, the method utilized to impose proportional loading throughout the deformation history, and the strain gradient plasticity material model are also presented in Section \ref{S:problemformulation}. The numerical results are presented and discussed in Section \ref{S:results}. Finally, the key results and conclusions of this work are summarized in Section \ref{sec:Conclusion}.

%%%%%%%%%%%%%%%%%%%%%%%%%%%%%%%%%%%%%%%%%%%%%%%%%%%%%%%%%%%%%%%%%%%%%%
\section{Problem formulation and modelling approach}
\label{S:problemformulation}

This work considers a limit load-type analysis to determine the critical stress level at which a given microstructure configuration loses load carrying capacity. Hence, an elastic-perfectly plastic material model is employed. The configuration of the unit cell and the simulation setup is described in Section \ref{sec:geometryandmesh}, while the approach to prescribe a constant value of stress triaxiality and Lode parameter is outlined in Section \ref{sec:numericalmethod}.

\subsection{Unit cell geometry and FE mesh}
\label{sec:geometryandmesh}

\begin{figure}[htbp!] 
\centerline{\includegraphics[width=3.5in]{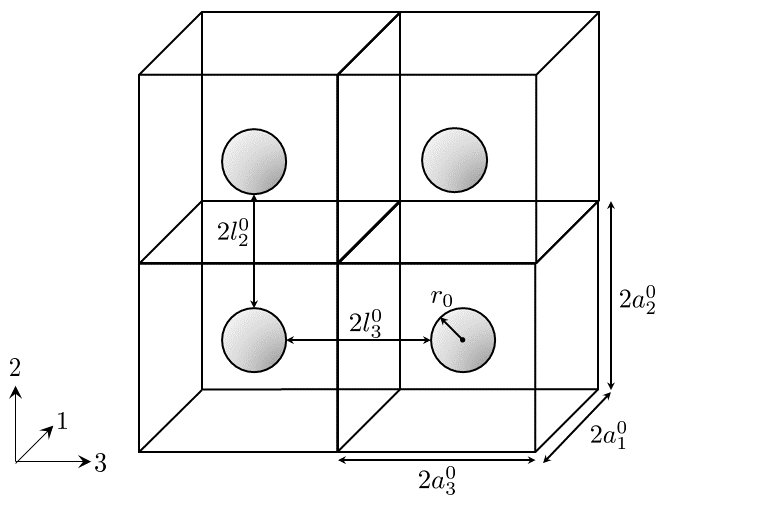}}
\caption{Schematic showing the periodic arrangement of voids in the $x_2$- and $x_3$-plane. The distribution along the $x_1$-direction is not shown for simplicity.}
\label{fig:cell}
\end{figure}

Three dimensional finite element calculations are carried out to model the response of an array of spherical voids with initial radius $r_0$, Fig.~\ref{fig:cell}. The unit cell has edge lengths $2a_i^0$ along the three coordinate axes, $x_i$ ($i=1,2,3$), and inter-void spacings thereby equal $2l_i^0=2a_i^0-2r_0$. Symmetry about three planes perpendicular to the coordinate axes implies that only $1/8$ of the unit cell needs to be modelled. 

\begin{figure}[htb!] 
\begin{overpic}[width=\linewidth]{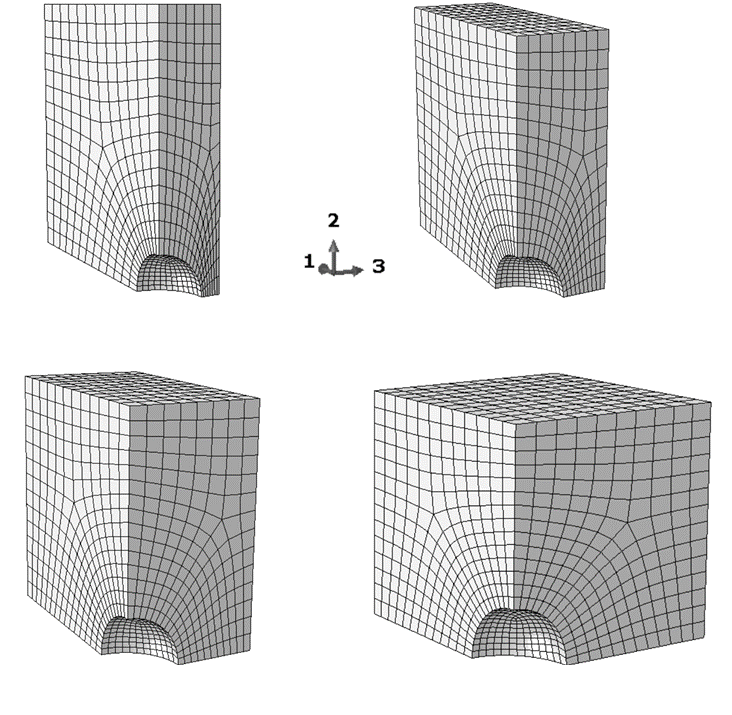}
\put(18,51){(a)}
\put(70,51){(b)}
\put(18,0){(c)}
\put(70,0){(d)}
\end{overpic}
\caption{Finite element meshes showing 1/8 of the unit cell with an initially spherical void of radius $r_0$ in the centre giving an initial void volume fraction of $f_0=0.01$ for (a) $\nicefrac{l_1^0}{r_0}=\nicefrac{l_2^0}{r_0}=5.06$; $\nicefrac{l_3^0}{r_0}=0.43$, (b) $\nicefrac{l_1^0}{r_0}=\nicefrac{l_2^0}{r_0}=4.21$; $\nicefrac{l_3^0}{r_0}=0.95$, (c) $\nicefrac{l_1^0}{r_0}=\nicefrac{l_2^0}{r_0}=3.58$; $\nicefrac{l_3^0}{r_0}=1.5$ and (d) $\nicefrac{l_1^0}{r_0}=\nicefrac{l_2^0}{r_0}=2.75$; $\nicefrac{l_3^0}{r_0}=2.75$. The number of elements ranges from 1896 for (a) to 2512 for (d).}    
\label{fig:meshes}
\end{figure}

For all unit cells considered, the initial void volume fraction is $f_0=0.01$, where $f_0=(4/3\pi r_0^3)/(8a^0_1 a^0_2 a^0_3)$. The initial void radius, $r_0$, is kept constant, while the cell dimensions are varied to achieve various initial inter-void spacings as in ref.~\cite{srivastava2013void}. The geometric parameters for the different cell dimensions are given in Table~\ref{tab:geometry}. For all unit cells, $a^0_1/r_0=a^0_2/r_0$. Finite element meshes for four unit cell configurations are shown in Fig.~\ref{fig:meshes}. The modelling setup does not account for softening due to void evolution since a small strain formulation is used. It is assumed that the unit cell represents the material condition immediately before failure, neglecting the deformation history leading to this state. Hence, model predictions for the loss of load carrying capacity signal the onset of localisation. The critical equivalent stress at the onset of localisation is recorded and reported in the results section. 

\begin{table}[htbp!]
\centering
\caption{Geometric parameters for the various unit cells considered for $f_0=0.01$. Based on \cite{srivastava2013void}.}
\label{tab:geometry}
\begin{tabular}{cccccc}
\hline
$a^0_1/r_0=a^0_2/r_0$ & $a^0_3/r_0$ & $l^0_1/r_0=l^0_2/r_0$ & $l^0_3/r_0$ \\
\hline       
$6.06$   & $1.43$        & $5.05$        & $0.43$        \\
$5.55$   & $1.70$        & $4.55$        & $0.70$        \\
$5.21$   & $1.94$        & $4.21$        & $0.95$        \\
$4.97$   & $2.12$        & $3.97$        & $1.12$        \\
$4.58$   & $2.50$        & $3.58$        & $1.50$        \\
$4.18$   & $3.00$        & $3.18$        & $2.00$        \\
$3.75$   & $3.75$        & $2.75$        & $2.75$        \\
\hline
\end{tabular}
\end{table}

\subsection{Numerical method}
\label{sec:numericalmethod}

The unit cells are subject to prescribed displacements and the boundary conditions applied to the faces of the cell are

\begin{align}
\label{eq:BC}
    u_1(a_1^0, x_2, x_3)=U_1(t),\quad T_2(a_1^0, x_2, x_3)=T_3(a_1^0, x_2, x_3)=0 \nonumber \\ 
    u_2(x_1, a_2^0, x_3)=U_2(t),\quad T_1(x_1, a_2^0, x_3)=T_3(x_1, a_2^0, x_3)=0 \nonumber \\
    u_3(x_1, x_2, a_3^0)=U_3(t),\quad T_1(x_1, x_2, a_3^0)=T_2(x_1, x_2, a_3^0)=0
\end{align}

\noindent The applied symmetry boundary conditions are

\begin{align}
\label{eq:BC_symm}
    u_1(0, x_2, x_3)=0, \quad T_2(0, x_2, x_3)=T_3(0, x_2, x_3)=0 \nonumber \\ 
    u_2(x_1, 0, x_3)=0, \quad T_1(x_1, 0, x_3)=T_3(x_1, 0, x_3)=0 \nonumber \\
    u_3(x_1, x_2, 0)=0, \quad T_1(x_1, x_2, 0)=T_2(x_1, x_2, 0)=0
\end{align}

In Eq. \eqref{eq:BC}, $U_1(t)$ is prescribed and the time history of the displacements $U_2(t)$ and $U_3(t)$ are determined such that a prescribed stress state is maintained. The loading direction is fixed in stress space by enforcing constant ratios between the normal stress components throughout the deformation history such that

\begin{align}
\label{eq:rho}
    \Sigma_{22} = \rho_2\Sigma_{11}, \qquad \Sigma_{33} = \rho_3\Sigma_{11}, 
\end{align}

\noindent where $\rho_2$ and $\rho_3$ are constants. The overall stress components $\Sigma_{ij}$ are found by volume averaging over all elements, such that: $\Sigma_{ij}=\int_V\sigma_{ij}\text{d}V/V$, where $V$ is the unit cell volume.

The overall effective stress, $\Sigma_e$, and the overall hydrostatic stress, $\Sigma_h$, are given by

\begin{align}
    \Sigma_e=\frac{1}{\sqrt{2}}\sqrt{(\Sigma_{11}-\Sigma_{22})^2+(\Sigma_{22}-\Sigma_{33})^2+(\Sigma_{33}-\Sigma_{11})^2}, \nonumber\\ \qquad \Sigma_h=\frac{1}{3}(\Sigma_{11}+\Sigma_{22}+\Sigma_{33})\nonumber,
\end{align}
which in terms of the relative stress ratios become
\begin{align}
    \Sigma_e=\Sigma_{11}\frac{1}{\sqrt{2}}\sqrt{(1-\rho_2)^2+(\rho_2-\rho_3)^2+(\rho_3-1)^2}, \\ \Sigma_h=\Sigma_{11}\frac{1}{3}(1+\rho_2+\rho_3).
\end{align}
The stress triaxiality, $T$, and the Lode parameter, $L$, are given by
\begin{align}
    T=\frac{\Sigma_h}{\Sigma_e}=\frac{\sqrt{2}}{3}\frac{1+\rho_2+\rho_3}{\sqrt{(1-\rho_2)^2+(\rho_2-\rho_3)^2+(\rho_3-1)^2}}
    \label{eq:T}
\end{align}
and
\begin{align}
    L=\frac{2\Sigma_{22}-\Sigma_{11}-\Sigma_{33}}{\Sigma_{11}-\Sigma_{33}}=\frac{2\rho_2-1-\rho_3}{1-\rho_3}.
    \label{eq:L}
\end{align}
The overall effective strain, $\overline{E}_e$, is given by
\begin{align}
    \overline{E}_e=\frac{\sqrt{2}}{3} \sqrt{(E_{11}-E_{22})^2+(E_{22}-E_{33})^2+(E_{33}-E_{11})^2}
\end{align}
where the strain components, $E_{ij}$, are found in a way analogous to the stress components. 

\subsubsection{Multiple point constraints}

The macroscopic normal stress components vary throughout the deformation, but the stress ratios are maintained in each increment of the simulation according to Eq.~\eqref{eq:rho}. This is achieved by creating multi point constraints through the user subroutine MPC in ABAQUS, which enables enforcing relationships between degrees of freedom in one or more nodes. 

\begin{figure}[htb!] 
\centering
\centerline{\includegraphics[width=2.5in]{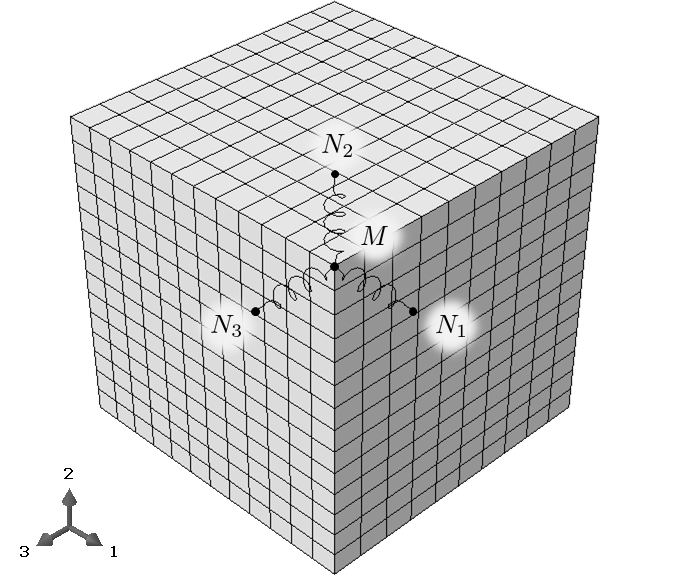}}
\caption{The spring elements for the multiple point constraints connected to one connector node, $M$, in the finite element mesh.}
\label{fig:MPC}
\end{figure}

Additional degrees of freedom are added to impose boundary conditions on all sides of the model while prescribing the stress ratios. Three dummy nodes, $N_i$, are created outside of the mesh and connected to one connector node, $M$, in the mesh as shown in Fig.~\ref{fig:MPC}. This connection is made through spring elements (SPRING2 elements from the ABAQUS element library). The displacement in the $x_1$-direction is then prescribed at the $N_1$-dummy node, while the displacements (in the $x_2$- and $x_3$-directions) corresponding to the desired stress triaxiality and Lode parameter are calculated and applied to the $N_2$- and $N_3$-dummy nodes. The displacement of the connector node, $M$, is coupled to the displacement of the nodes located at $(a_1^0,x_2,x_3)$, $(x_1,a_2^0,x_3)$ and $(x_1,x_2,a_3^0)$ in the direction of the respective face normals. In this way, the displacement of the dummy nodes, $N_i$, is linked to the unit cell. 

The displacement of the dummy nodes, $N_i$, is related to the forces, $F_i$, at the faces of the unit cell through

\begin{align}
   F_i=k_i(u_i^{N_i}-u_i^M) \quad \text{with} \quad i=1,2,3,
\label{eq:springs}
\end{align}

\noindent and $k_i$ being the spring element constants given by $k_i=E(\nicefrac{A_i}{a_i^0})\times10^{-1}$, where the factor of $10^{-1}$ is introduced to stabilise the numerical solution, following Ref.~\cite{tekoglu2014representative}. The forces, $F_i$, are the resultant of all traction across the corresponding surface and relates to the macroscopic stresses through

\begin{align}
    \Sigma_{11}=\frac{F_1}{A_1}, \qquad A_1=a_2^0a_3^0 \nonumber \\
    \Sigma_{22}=\frac{F_2}{A_2}, \qquad A_2=a_1^0a_3^0 \nonumber \\
    \Sigma_{33}=\frac{F_3}{A_3}, \qquad A_3=a_1^0a_2^0
\label{eq:stress}
\end{align}

\noindent where $A_i$ is the area over which the forces act. Combining Eqs.~\eqref{eq:rho}, \eqref{eq:springs}, and \eqref{eq:stress}, gives the dummy node displacements

\begin{align}
    \rho_2=\frac{\Sigma_{22}}{\Sigma_{11}}=\text{const.} \Rightarrow u_2^{N_2}=u_2^M+\rho_2\frac{A_2}{A_1}\frac{k_1}{k_2}(u_1^{N_1}-u_1^M) \nonumber \\
    \rho_3=\frac{\Sigma_{33}}{\Sigma_{11}}=\text{const.} \Rightarrow u_3^{N_3}=u_3^M+\rho_3\frac{A_3}{A_1}\frac{k_1}{k_3}(u_1^{N_1}-u_1^M),
\end{align}

\noindent where $\rho_2$ and $\rho_3$ are input values for the stress ratio, $u_i^{N_j}$ is the displacement of dummy node $j$ in the direction of $x_i$, $u_i^M$ is the displacement in $x_i$-direction of the connector node $M$, $A_i$ are areas from Eq.~\eqref{eq:stress}, and $k_i$ are the spring element constants. Another relevant procedure for imposing multiple point constraints without spring elements can be found in \cite{liu2016void}.

The calculations are carried out for three values of Lode parameter, $L=-1, 0$, and $1$. The Lode parameter values $L=-1$ ($\Sigma_{11}>\Sigma_{22}=\Sigma_{33}$) and $L=1$ ($\Sigma_{11}=\Sigma_{22}>\Sigma_{33}$) correspond to overall axisymmetric stress states, while $L=0$ ($\Sigma_{11}>\Sigma_{22}>\Sigma_{33}$) correspond to an overall state of shear plus hydrostatic stress. For each value of Lode parameter, three triaxialities are considered, $T=1, 2$, and $3$. The values for $\rho_2$ and $\rho_3$ to achieve these stress states are given in Table~\ref{tab:input}.

\begin{table}[htb!]
\caption{Input parameters determining the prescribed stress state.}
\label{tab:input}
\begin{tabular}{m{0.1\textwidth}m{0.1\textwidth}m{0.1\textwidth}m{0.1\textwidth}}
\hline
L & T & $\rho_2$ & $\rho_3$  \\
\hline
\multirow{4}{*}{-1} & $\frac{1+2\rho_2}{3(1-\rho_2)}$ & $\rho_3$ & $\frac{1+2\rho_2}{3(1-\rho_2)}$ \\
                    & 1 & 0.4 & 0.4    \\
                    & 2 & 0.625 & 0.625  \\
                    & 3 & 0.727273 & 0.727273  \\
\hline
\multirow{4}{*}{0} & $\frac{1+\rho_2}{\sqrt{3}(1-\rho_2)}$ & $\frac{1+\rho_3}{2}$ & $\frac{\sqrt{3}T-1}{\sqrt{3}T+1}$ \\
                   & 1 & 0.634 & 0.268 \\
                   & 2 & 0.776 & 0.552 \\
                   & 3 & 0.8386 & 0.6772 \\
\hline
\multirow{4}{*}{1} & $\frac{2+\rho_2}{3(1-\rho_2)}$  &  1 &  $\frac{3T-2}{3T+1}$ \\
                   & 1 & 1 & 0.25 \\
                   & 2 & 1 & 0.57 \\
                   & 3 & 1 & 0.70 \\
\hline
\end{tabular}
\end{table}

The calculations were carried out using the commercial finite element code ABAQUS with the gradient theory applied to the matrix material through a UEL subroutine. The reader is referred to Ref. \cite{martinez2019} for further details on the implementation. The calculations use 20-node user defined elements. The number of elements in the finite element meshes is varied from a minimum of 1896 to a maximum of 2512 elements, Fig.~\ref{fig:meshes}.

\subsection{Material model: Strain gradient plasticity}
\label{sec:gradientmodel}

The gradient enhanced constitutive model employed is based on the visco-plastic strain gradient plasticity theory proposed by Gudmundson \cite{gudmundson2004} in the context of the mathematical formulation in terms of minimum principles proposed by Fleck and Willis \cite{fleck2009}. For the dissipative version considered, the theory accounts for internal elastic energy storage due to elastic strain and dissipation due to the plastic strain rate, $\dot{\varepsilon}^p_{ij}$, and its spatial gradient, $\dot{\varepsilon}^p_{ij,k}$. Contributions from plastic strain gradients to free energy is ignored. The Principle of Virtual Work (PVW) in Cartesian components is expressed by

\begin{multline}
    \int_{V}\left(\sigma_{ij}\delta\dot{\varepsilon}_{ij}+(q_{ij}-s_{ij})\delta\dot{\varepsilon}^p_{ij}+\tau_{ijk}\delta\dot{\varepsilon}^p_{ij,k}\right)\text{d}V= \\ \int_{S}\left(T_i\delta\dot{u}_i+t_{ij}\delta\dot{\varepsilon}^p_{ij}\right)\text{d}S
\end{multline}

\noindent where $\sigma_{ij}$ and $s_{ij}=\sigma_{ij}-\frac{1}{3}\delta_{ij}\sigma_{kk}$ are the Cauchy stress tensor and the stress deviator, respectively. The micro-stress, $q_{ij}$, is work conjugate to the plastic strain rate, $\dot{\varepsilon}^p_{ij}$, and $\tau_{ijk}$ is a higher order stress, work conjugate to the plastic strain rate gradient, $\dot{\varepsilon}^p_{ij,k}$. The right hand side of the PVW includes the conventional traction, $T_{i}=\sigma_{ij}n_j$ work conjugate to the boundary displacement rate, $\dot u_i$, and the higher order traction, $t_{ij}=\tau_{ijk}n_k$, work conjugate to the plastic strain rate, $\dot{\varepsilon}_{ij}^p$. Here, the outward unit normal to the surface $S$ is $n_i$.  Balance laws for the stress quantities are given by

\begin{gather}
    \sigma_{ij,j}=0 \quad \text{and} \quad q_{ij}-s_{ij}-\tau_{ijk,k}=0
\end{gather}

\noindent where, the first set of equations is the conventional equilibrium equations in the absence of body forces, and the second set is the higher order equilibrium equations. The higher order boundary conditions are imposed such that the void surface is higher order traction free, while symmetry conditions are imposed at the exterior of the cell through $\varepsilon_{12}=0$.

\subsection{Constitutive equations}

The rate-dependent visco-plastic formulation employs a potential to account for plastic dissipation as follows

\begin{equation}
  \Phi\left[\dot{E}^p,E^p\right]=\int_{0}^{\dot{E}^p}\sigma_c\left[\dot{E}^{p'},E^p\right]\text{d}\dot{E}^{p'}
\end{equation}

Here, $\sigma_c$ is the gradient enhanced effective stress, related to the current matrix flow stress through $\sigma_c=\sigma_F[E^p]\left(\frac{\dot{E}^p}{\dot{\varepsilon}_0}\right)^m$, with $\dot{\varepsilon}_0$ denoting the reference strain rate, and $m$ denoting the rate-sensitivity exponent. The material in this work does not undergo strain hardening, making $\sigma_F$ independent of $\dot{E}^p$ and equal to the material yield stress $\Sigma_0$. The viscoplastic law is implemented following the algorithm presented in Ref. \cite{FuentesVisco} to efficiently approach the rate-independent limit. A gradient enhanced effective plastic strain rate is given by

\begin{equation}
\label{eq:dotE}
    \left(\dot{E}^p\right)^2 = \frac{2}{3}\dot{\varepsilon}^p_{ij}\dot{\varepsilon}^p_{ij}+L_D^2\dot{\varepsilon}^p_{ij,k}\dot{\varepsilon}^p_{ij,k}
\end{equation}

\noindent and the associated work conjugate gradient enhanced effective stress by

\begin{equation}
\label{eq:sigmac}
    \sigma^2_c=\frac{3}{2}q_{ij}q_{ij}+\frac{1}{L_D^2}\tau_{ijk}\tau_{ijk}.
\end{equation}

\noindent Here, $L_D$ is a dissipative constitutive length parameter that enters for dimensional consistency. The superscript $D$ refers to dissipative quantities, and the dissipative stress quantities are given by

\begin{equation}
\label{eq:disstress}
    q^D_{ij}=\frac{2}{3}\sigma_c\frac{\dot{\varepsilon}^p_{ij}}{\dot{E}^p}, \,\,\,\,\,\,\,\, \tau^D_{ijk}=L_D^2\sigma_c\frac{\dot{\varepsilon}^p_{ij,k}}{\dot{E}^p}.
\end{equation}

The dissipative length parameter controls the strengthening size effect with an increase in the dissipative length parameter giving an increase in the apparent yield stress in the presence of strain gradients, see~\cite{martinez2016finite,voyiadjis2012thermo}. This work is a limit load analysis, which, by definition, is done to determine the overall yield criterion for a given, specific configuration. Limit load analyses normally idealise materials as perfectly plastic. To avoid strain hardening from the energetic gradient contributions, the energetic length parameter, $L_E$, has been set to zero in this work, and, consequently, the corresponding energetic quantities are omitted.

%%%%%%%%%%%%%%%%%%%%%%%%%%%%%%%%%%%%%%%%%%%%%%%%%%%%%%%%%%%%%%%%%%%%%%
\section{Numerical results and discussion}
\label{S:results}

Throughout, the following material parameters are used; $\Sigma_{0}/E=0.001$, $\nu=0.3$ and $m=0.01$, where $\Sigma_{0}$ is the yield stress, $E$ is Young's modulus, $\nu$ is the Poisson ratio, and $m$ is the strain rate sensitivity exponent. The value of $m$ is considered sufficiently small for the results to approximate a rate-independent material response. The influence of the Lode parameter, $L$, the stress triaxiality, $T$, and the normalised length parameter, $L_D/r_0$, is studied. The effect of the inter-void ligament size is discussed in combination with the other parameters, $L$, $T$, and $L_D/r_0$. 

\subsection{Critical equivalent stress at localisation}
\label{sec:criticalstress}

\begin{figure*}[htb!]
\begin{minipage}[c]{0.45\textwidth}
    \begin{overpic}[width=\textwidth]{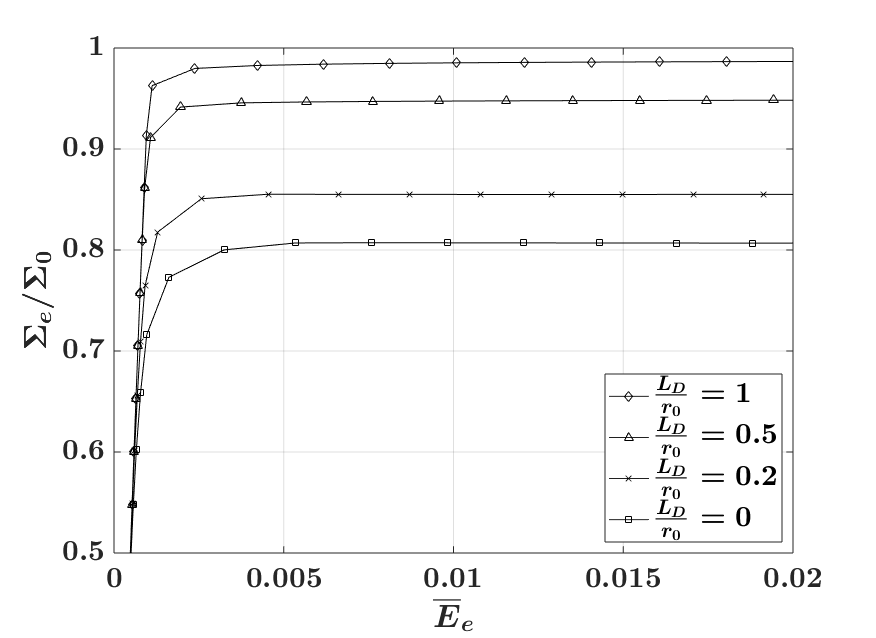}
    \put(17,12){\textbf{$L=-1$}}
    \end{overpic}
\subcaption{}    
\end{minipage}
\hspace{0.2cm}  
\begin{minipage}[c]{0.45\textwidth}
    \begin{overpic}[width=\textwidth]{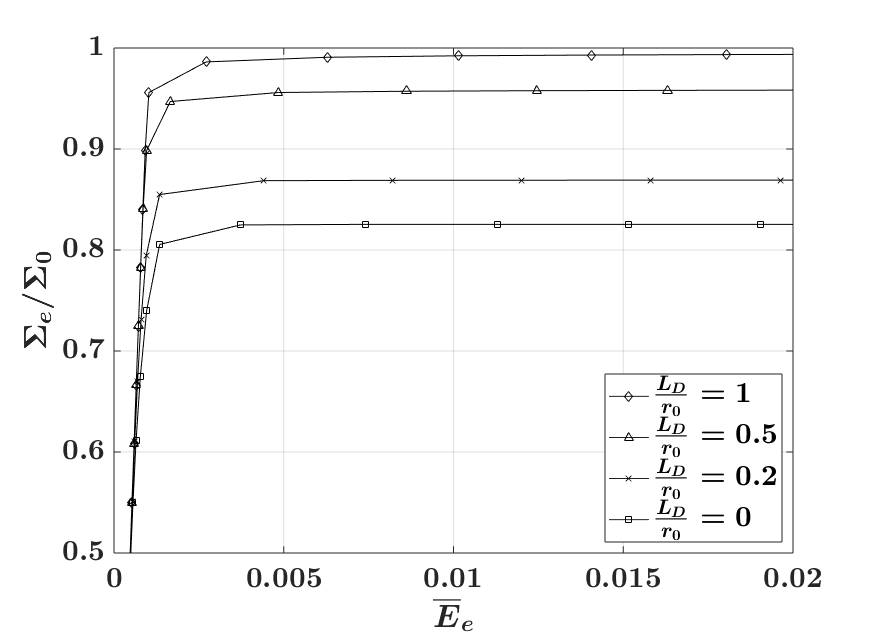}
    \put(17,12){\textbf{$L=1$}}
    \end{overpic}
\subcaption{}    
\end{minipage}
\caption{Equivalent stress-strain curve for an inter-void ligament size of $l_3/r_0=1.5$ under loading conditions giving Lode parameters of (a) $L=-1$ and (b) $L=1$ and a triaxiality of $T=3$.}
\label{fig:stress-strain}
\end{figure*}

Figure~\ref{fig:stress-strain} presents the equivalent stress-strain curves for two distinct Lode parameters, $L=-1$ and $1$, but for a fixed stress triaxiality, $T=3$, and a fixed inter-void ligament size of $l_3/r_0=1.5$. The equivalent stress-strain curves are depicted for three length parameters, being, $L_D/r_0=0.2, 0.5$, and $1$ as well as for the conventional limit where $L_D/r_0=0$. 

The material response shows a clear effect of plastic strain gradients, such that the larger the length parameter, the higher the equivalent stress level. This means that an increase in the stress level is obtained when down-scaling the microstructure and, thus, yielding of the material is delayed due to increasing strain gradient strengthening. The critical equivalent stress, $\Sigma_e^c/\Sigma_0$, signaling localisation (and coalescence) is taken to be at the plateau of the equivalent stress-strain curve. Several ways exists to establish a coalescence criteria based on either critical stress or strain. The method employed in this work is inspired by the work of~\cite{tekoglu2012criterion}. The critical stresses have been extracted from the end of the equivalent stress-strain curves (as shown in Figures 4a and b). Taking the example of Fig.~\ref{fig:stress-strain}a, the conventional material ($L_D/r_0=0$), the difference between the equivalent stress at $E_e=0.02$ and $0.01$ is less than $0.04\%$. %\textcolor{red}{For coalescence studies using finite strain formulations, a commonly used coalescence criterion is the overall critical strain. For the adopted small strain formulation, however, the critical equivalent stress is deemed an appropriate coalescence criterion for a perfectly plastic material. This is a simple and effective measure allowing for straight-forward determination of the critical coalescence stress.} 

\subsection{Conventional material: Effect of the inter-void ligament size}
\label{s:Lodeparameter}

The conventional limit, $L_D/r_0=0$, is considered to set the scene for the study of material size effects. The focus here is the effect of inter-void ligament size on the critical stress at localisation under various loading conditions.

\begin{figure}[htb!] 
\centerline{\includegraphics[width=3.34in]{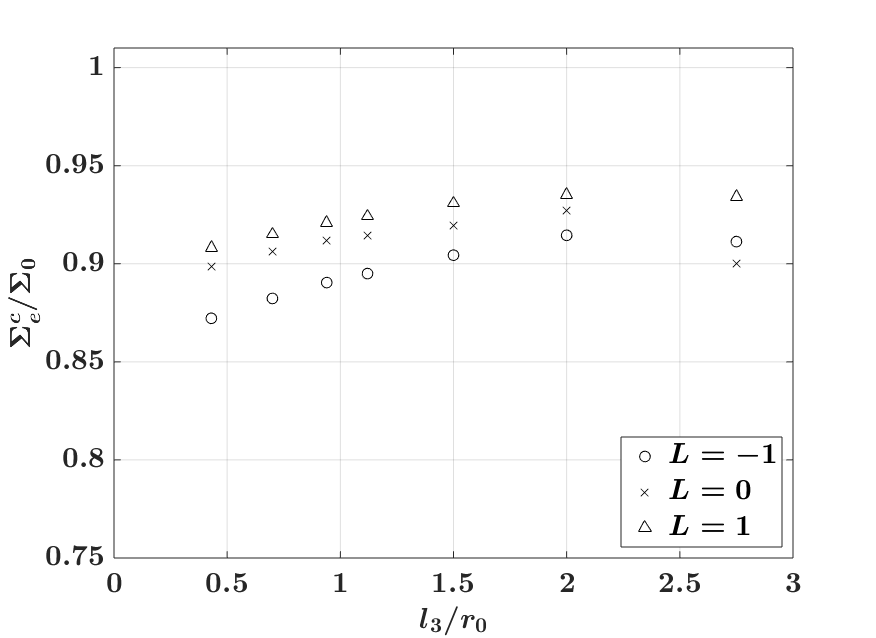}}
\caption{Critical equivalent stress vs. normalized inter-void ligament size for three values of the Lode parameter with $T=2$ and $L_D/r_0=0$.}
\label{fig:Lode}
\end{figure}

First, three values of the Lode parameter are considered, $L=-1, 0$, and $1$, for a fixed stress triaxiality, $T=2$. Figure~\ref{fig:Lode} shows the critical equivalent stress as a function of the inter-void ligament size. For the six smallest inter-void ligaments, the critical equivalent stress is seen to increase when the inter-void ligament becomes bigger irrespective of the value of the Lode parameter. The increase in the critical stress ties to localisation occurring more easily in small inter-void ligaments lowering the load carrying capacity of the unit cell. As the inter-void ligament size increases, the $l_3$-ligament can sustain a higher stress level before localisation, leading to an increase in critical equivalent stress. Also, for the six smallest inter-void ligaments, an increase in the critical stress is found with increasing Lode parameter values. Thus, the lowest critical equivalent stress is found for $L=-1$. The dependence on the Lode parameter can be rationalised by considering the imposed stress state. In comparison to the other cases, the relative stress component, $\rho_3$, is the largest when $L=-1$ (see Tab.~\ref{tab:input}), and localisation is therefore expected in the $l_3$-ligament at a lower overall deformation. In contrast, the $\rho_3$ takes the lowest value for $L=1$, resulting in delayed localisation and the highest critical equivalent stress obtained. In Ref. \cite{srivastava2013void}, void coalescence was found to occur along the ligament with the smallest applied stress for $L>-1$. For all Lode parameters, the relative stress component in the $l_3$-ligament will be smallest as $\rho_3$ is always the lowest stress ratio. For $L=-1$, coalescence occurs in the direction of the smallest inter-void ligament size. This corresponds to the $l_3$-ligament for all geometries except when $l_3/r_0=2.75$ as this is a perfect cube, Table~\ref{tab:geometry}. 

\begin{figure}[htb!]
\centering
\begin{overpic}[width=\linewidth]{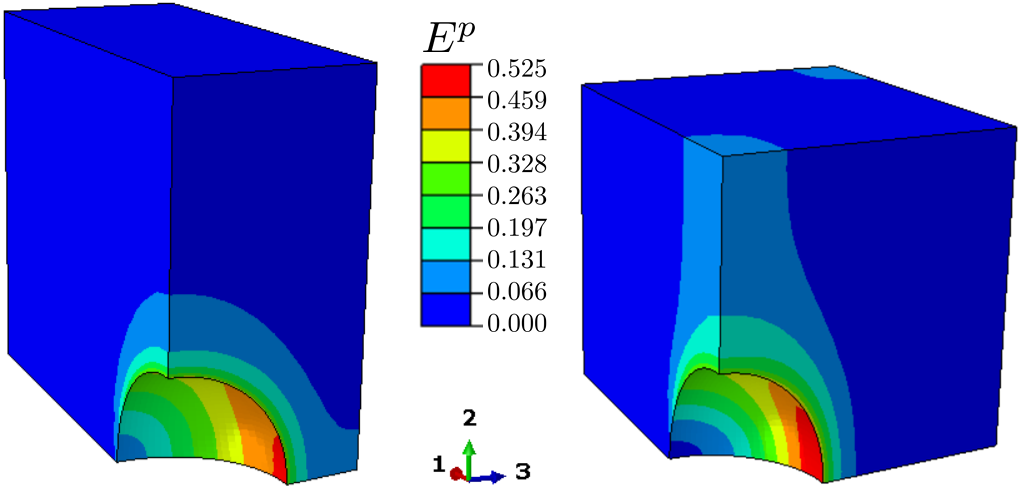}
\end{overpic}
\caption{Distribution of effective plastic strain for $L=0$, $T=2$, $L_D/r_0=0$, for $l_3/r_0=1.5$ to the left and $l_3/r_0=2.75$ to the right at a macroscopic effective strain of $\overline{E}_e=0.03$. For $l_3/r_0=1.5$, localisation is favoured in the smallest ligament, $l_3$. For the cubic unit cell, however, there is no bias towards any of the ligaments and deformation localises along $\approx 45^\circ$, i.e. across the diagonal.}
\label{fig:contour_Case5+7}
\end{figure}

\begin{figure}[htbp!]
\centering
\begin{overpic}[width=0.5\textwidth]{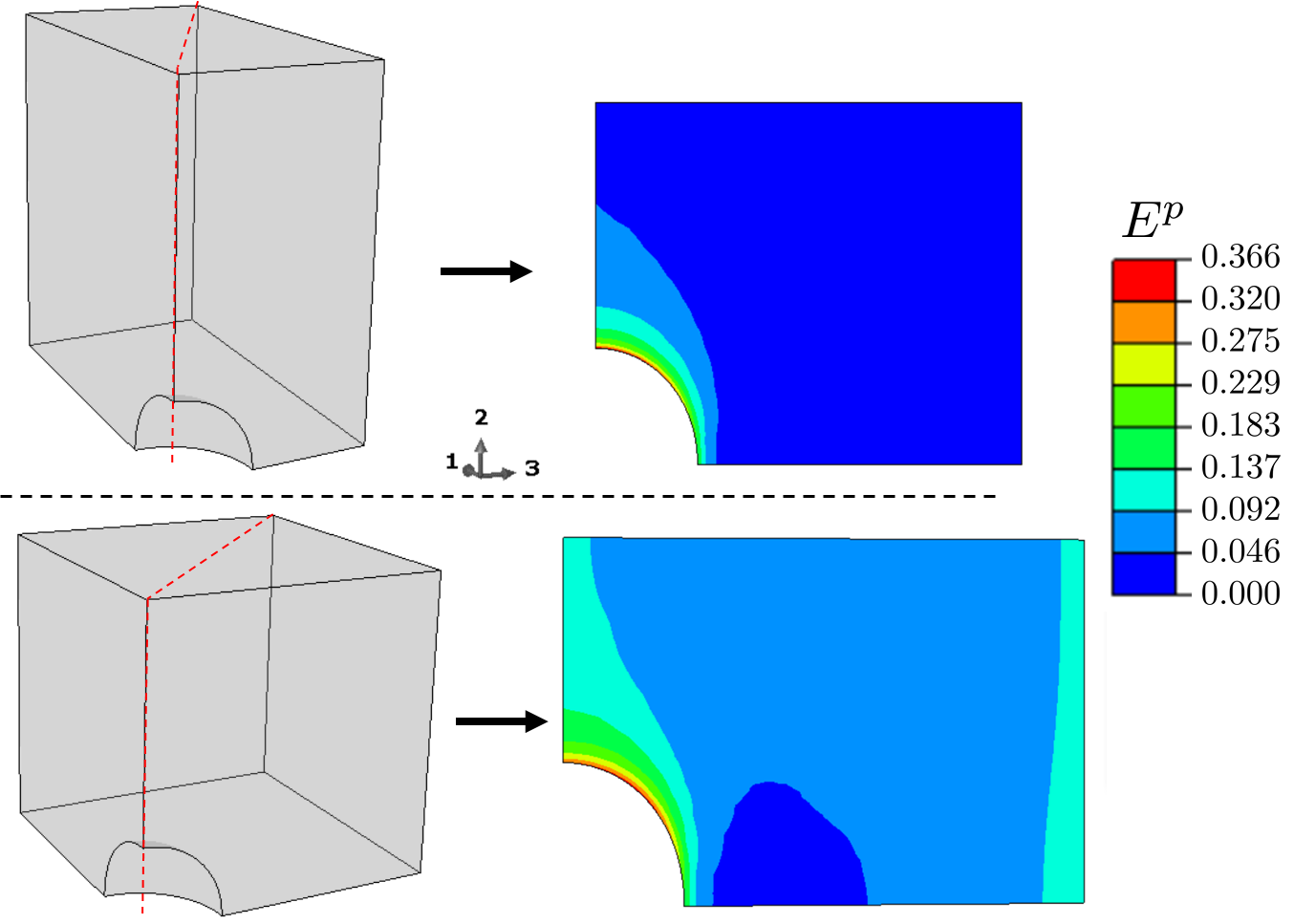}
\put(0,35){(a)} \put(0,0){(b)}
\end{overpic}
\caption{Distribution of effective plastic strain along a cut from corner to corner for $L=0$, $T=2$, $L_D/r_0=0$ at $\overline{E}_e=0.03$ for two geometries: (a) $l_3/r_0=1.5$ and (b) $l_3/r_0=2.75$. In (a) plasticity has not localized along $45^\circ$ and for this geometry localisation is favoured in the smaller ligament, while (b) shows that a band, indicated by the dotted line, has formed at a $45^o$ angle to the main loading axis (the $x_1$-axis), thus lowering the critical effective stress.}
\label{fig:Shearband}
\end{figure}

There is, however, a shift in the localisation pattern when the $l_3$-ligament becomes sufficiently wide, for example, a drop in the coalescence stress is found for $l_3/r_0=2.75$ for $L=0$. The load carrying capacity of the material increases when the distribution of voids diverge from a regular array, i.e. when $l_3/r_0 \neq 2.75$. However, the load carrying capacity will decrease if the arrangement of the voids is such that the inter-void ligament size, in any direction, is too small (e.g. $l_3/r_0 <  2$). At the configuration with $l_3/r_0=2.75$, there is no bias towards the $l_3$-ligament since the unit cell takes a cubic shape. The shift in the localisation is especially prominent for $L=0$ (a state of combined hydrostatic tension and shear) where plastic flow localises at $\approx 45^\circ$ across the cubic unit cell leading to an early loss of load carrying capacity. The shift in the localisation is demonstrated by depicting the contours of the effective plastic strain for two distinct unit cells ($l_3/r_0=1.5$ and $2.75$) subjected to $L=0$ and $T=2$ in Figs.~\ref{fig:contour_Case5+7} and \ref{fig:Shearband}. The material response remains conventional such that $L_D/r_0=0$, and the loading conditions are described by $L=0$ and $T=2$. For the conventional material, the second term of Eq.~\eqref{eq:dotE} is zero ($L_D=0$) and the term \textit{gradient enhanced effective plastic strain} refers to the time integration of only the first term of Eq. \eqref{eq:dotE}. For the elongated unit cell ($l_3/r_0=1.5$), localisation is seen to occur in the smallest ligament, $l_3$, whereas localisation is seen to occur along two corners of the cubic unit cell ($l_3/r_0=2.75$), indicating that the deformation is localised along $\approx 45^\circ$ i.e. across the diagonal. Figure~\ref{fig:contour_Case5+7} shows the contour of equivalent plastic strain across the faces of the cell, while Fig.~\ref{fig:Shearband}(a) and (b) show the contour of the effective plastic strain in the diagonal cross-section of both unit cells at an overall effective strain of $\overline{E}_e=0.03$. By comparing the two contours it is seen that plastic flow is observed across the entire cross-section indicating localisation at $\approx 45^\circ$ for the cubic model, $l_3/r_0=2.75$. In contrast, the plastic flow is constricted for $l_3/r_0=1.5$ shown top right in Fig.~\ref{fig:Shearband}.

\begin{figure}[htb!]
\centering
    \includegraphics[width=3.34in]{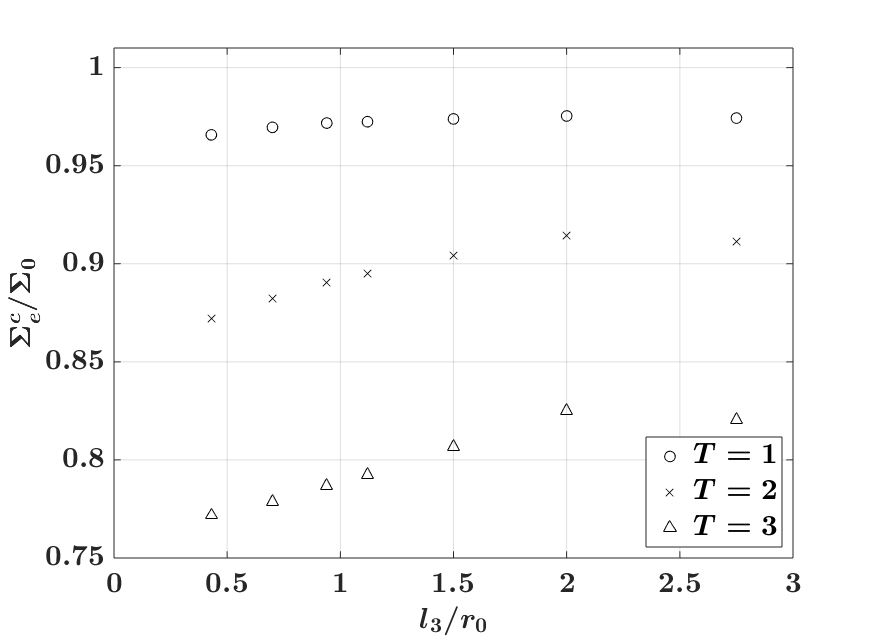}
    \caption{Critical equivalent stress vs. normalized inter-void ligament size for three values of the stress triaxialities with $L=-1$ and $L_D/r_0=0$.}
    \label{fig:triaxiality}
\end{figure}

Next, the effect of stress triaxiality is considered. In Fig.~ \ref{fig:triaxiality}, the critical stress for a conventional material, $L_D/r_0=0$, is shown as a function of the inter-void ligament size for $T=1, 2,$ and $3$ for a fixed value of the Lode parameter, $L=-1$. In the conventional limit, a high level of stress triaxiality yields low critical stress for all ligament sizes considered. The reason being that a high stress triaxiality corresponds to higher relative stress components, $\rho_2$ and $\rho_3$. Figure~\ref{fig:triaxiality} shows little effect of ligament size on the critical equivalent stress for the low value of triaxiality. For $T=1$, the relative stress transverse to the main loading direction is insufficient to invoke localisation in the inter-void ligament and the effect of the ligament size will be limited. The cell instead undergoes macroscopic localisation and, consequently, does not exhibit a profound dependence on the inter-void ligament size. This is in line with results presented in Ref. \cite{tekouglu2015localization}, where $T=1$ has been found to be the limit below which the onset of macroscopic localisation is essentially simultaneous with void coalescence. The results for $T=2$ and $T=3$ in Fig.~\ref{fig:triaxiality} show that the critical equivalent stress is dependent on inter-void ligament size. 

\begin{figure}[htb!]
\centering
    \includegraphics[width=2.75in]{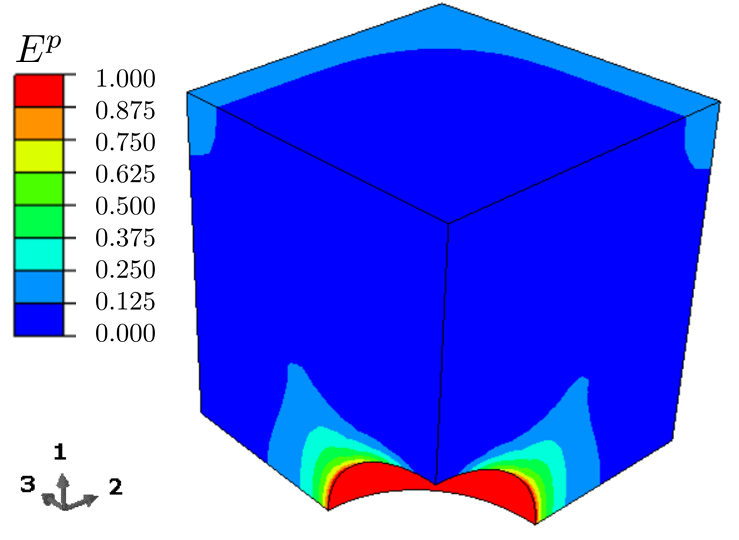}
    \caption{Bands of plastic flow in the cubic unit cell ($l_3/r_0=2.75$) and $L_D/r_0=0$ at an overall equivalent strain of $\overline{E}_e=0.08$. The loading conditions applied to give an axisymmetric stress state with $L=-1$ and $T=2$. Note the rotated coordinate system to show the symmetry of the plastic flow given by the cubic unit cell and $\rho_2=\rho_3$ for $L=-1$.}
    \label{fig:triaxiality_symmetry}
\end{figure}

A small drop in the critical equivalent stress is seen to occur for the largest ligament for all values of triaxiality in Fig.~\ref{fig:triaxiality}. The effect is most prominent for the highest triaxiality, $T=3$. Figure \ref{fig:triaxiality_symmetry} shows the contour gradient enhanced effective plastic strain of the cubic cell ($l_3/r_0=2.75$) at an effective stress of $\overline{E}_e=0.08$. At a sufficiently large strain, plasticity is seen to initiate at the corner opposite to the void. Due to the symmetry of both the loading condition ($\Sigma_2=\Sigma_3$ for $L=-1$) and the unit cell, bands of plastic deformation are observed to stretch across the $x_1-x_2$ and $x_1-x_3$ faces, ultimately lowering the coalescence stress giving the drop as seen in Fig.~\ref{fig:triaxiality}.

\subsection{Gradient enriched material: Effect of the inter-void ligament size}
\label{s:lengthscaleparamter}

\begin{figure*}
\centering
\begin{overpic}[width=\textwidth]{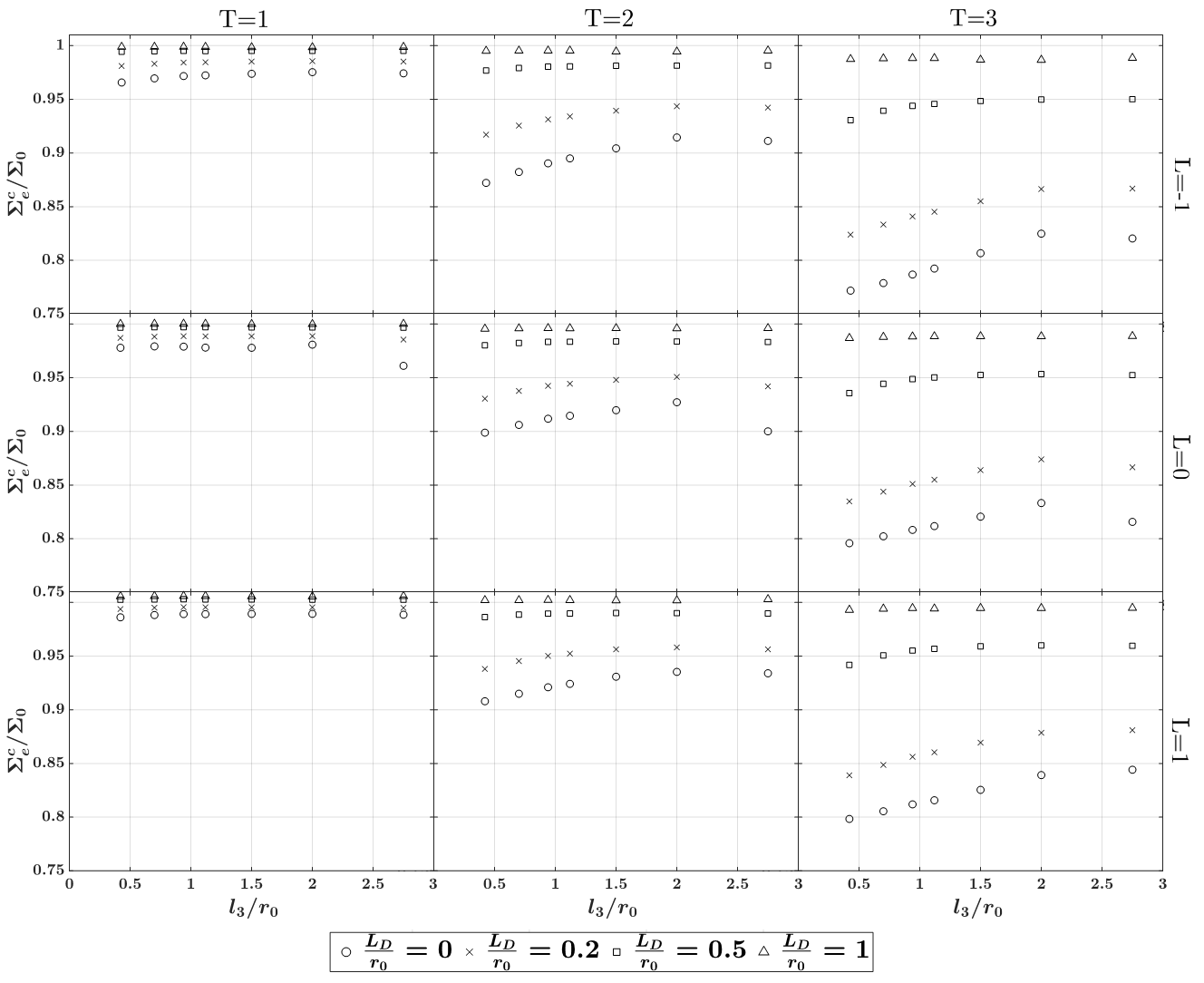}
\put(6.2,56.5){(a)} \put(37,56.5){(b)} \put(67,56.5){(c)}
\put(6.2,33){(d)} \put(37,33){(e)} \put(67,33){(f)}
\put(6.2,10){(g)} \put(37,10){(h)} \put(67,10){(i)}
\end{overpic}
\caption{The critical equivalent stress as a function of the smallest inter-void ligament size. Three values of the Lode parameter are considered, $L=-1, 0$, and $1$. For each Lode parameter, three values of the stress triaxiality are considered, $T=1, 2$, and $3$. Throughout, the parameters $\Sigma_0/E=0.001$, $\nu=0.3$ and $m=0.01$ are used. The initial void volume fraction is, $f_0=0.01$. The length parameter that enters through the gradient plasticity theory is $L_D/r_0=0.2, 0.5$ and $1$. A conventional material is modelled with $L_D/r_0=0$ and used as a reference.}
\label{fig:SigC_l3}
\end{figure*}
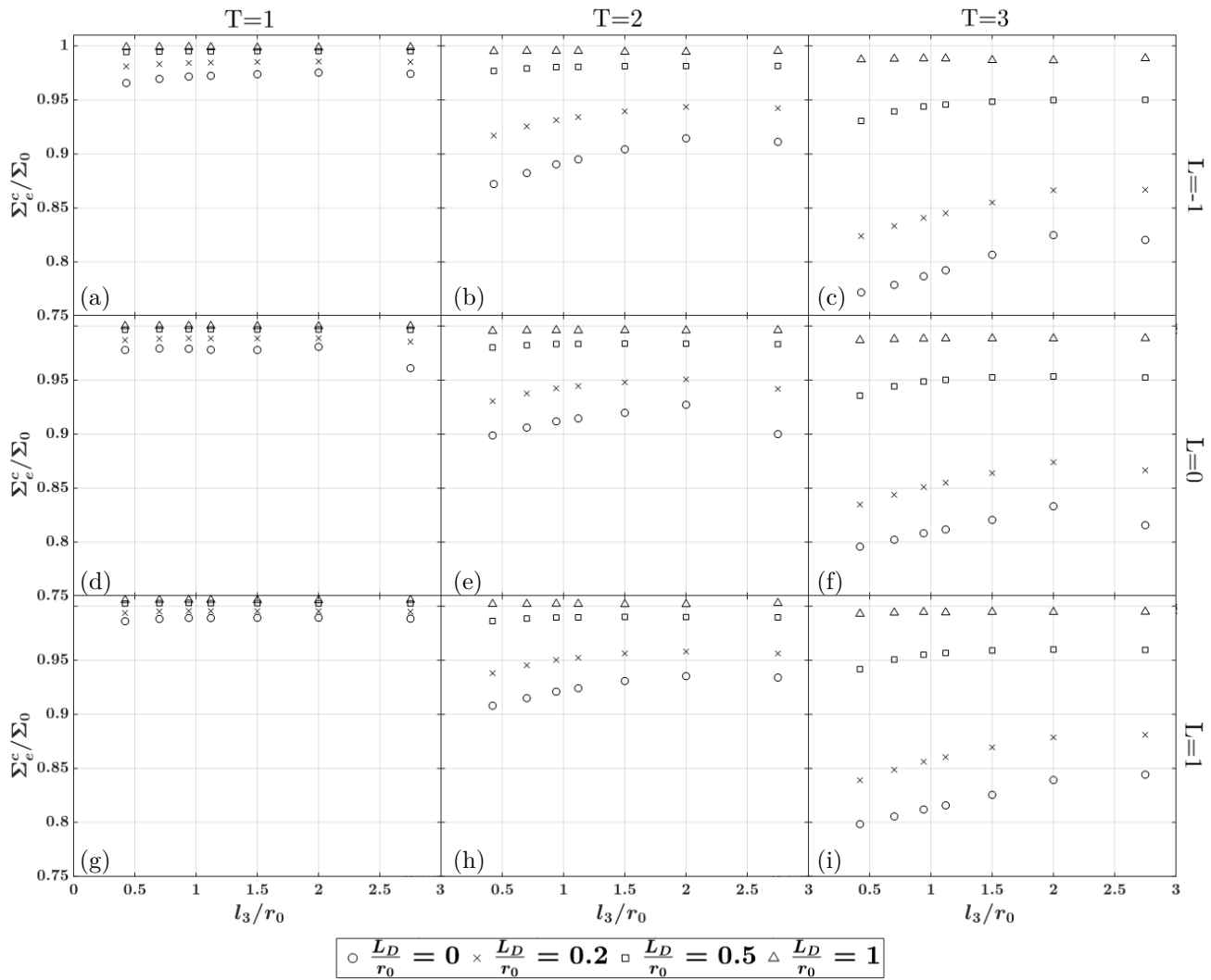

The effect of gradient strengthening in the matrix material is introduced through the length parameter $L_D$ (see Section~\ref{sec:gradientmodel}). One can imagine down-scaling the microstructure when increasing the value of $L_D/r_0$. Three values of the length parameter, $L_D/r_0=0.2, 0.5$, and $1$, are considered in the following for all combinations of the Lode parameter, $L=-1, 0, 1$, and stress triaxiality, $T=1, 2, 3$. Figure~\ref{fig:SigC_l3} shows the critical effective stress, $\Sigma_e^c/\Sigma_0$, as a function of the inter-void ligament size, $l_3/r_0$, for all combinations. The results obtained for a conventional material, $L_D/r_0=0$ are presented as a reference (see Section~\ref{s:Lodeparameter}).

The general observation is that the critical stress at localisation increases with the magnitude of the length parameter (down-scaling the microstructure). However, the critical stress has a natural upper bound where the gradient strengthening is so severe that the entire matrix material yields. At such large values of the length parameter, the effect of the Lode parameter, stress triaxiality, and inter-void ligament size vanish and the critical equivalent stress is identical for all combinations of geometry and loading condition. The threshold value is evident from Figs.~\ref{fig:SigC_l3}(a)-(i).

\begin{figure}
\centering
\begin{overpic}[width=0.5\textwidth]{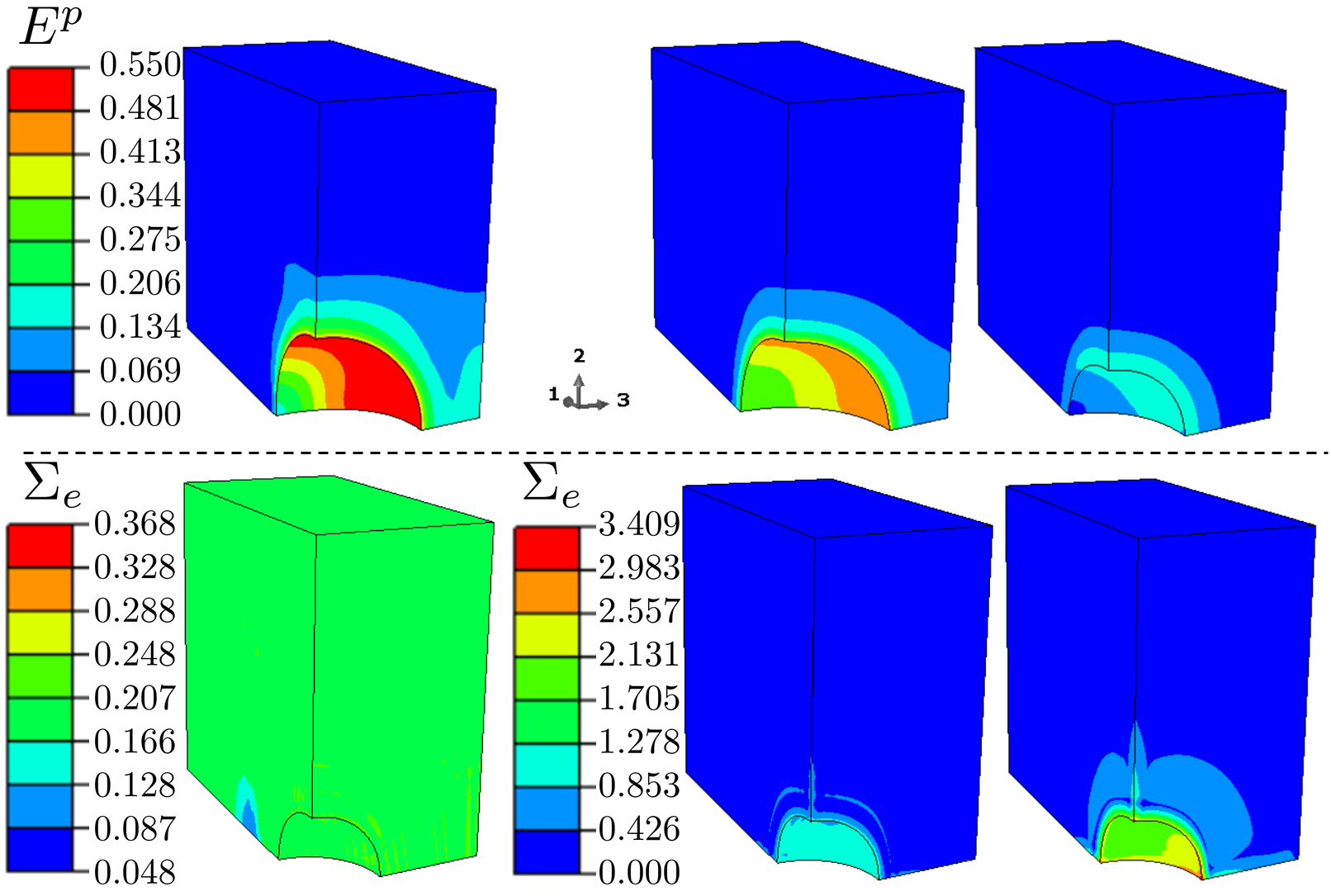}
\put(18,67){(a)} \put(55,67){(b)} \put(82,67){(c)}
\put(18,-2){(d)} \put(55,-2){(e)} \put(82,-2){(f)}
\end{overpic}
\vspace*{-3mm}
\caption{Distribution of gradient enhanced effective plastic strain for $L=-1$, $T=3$, $l_3/r_0=1.5$ for (a) the conventional material, $L_D/r_0=0$, (b) $L_D/r_0=0.2$ and (c) $L_D/r_0=0.5$ at a macroscopic effective strain of $\overline{E}_e=0.02$. The effective stress, $\Sigma_e$, for the same configuration is shown in the cells at the bottom, also here with (d) the conventional material, (e) $L_D/r_0=0.2$ and (f) $L_D/r_0=0.5$.}
\label{fig:contour_lengthscale}
\end{figure}

Figure~\ref{fig:contour_lengthscale}(a)-(c) show how the length parameter affects the plastic flow in the unit cells by comparing contours of the gradient enhanced plastic equivalent strain, $E^p$, (see Eq.~\eqref{eq:dotE}) for a ligament size of $l_3/r_0=1.5$ subject to $L=-1$ and $T=3$. The contours are extracted at an overall equivalent strain of $\overline{E}_e=0.02$. Figure~\ref{fig:contour_lengthscale}(a) displays the conventional material response where localisation occurs in the $l_3$-ligament. At the same level of the overall deformation, a significantly lower effective plastic strain is observed in Fig.~\ref{fig:contour_lengthscale}(b) and (c) when increasing the length parameter. For $L_D/r_0=0.2$, Fig.~\ref{fig:contour_lengthscale}(b), some plasticity is seen to develop in the $l_3$-ligament, but far less than in the conventional case, while the plasticity has barely initiated at this level of the deformation for $L_D/r_0=0.5$ Fig.~\ref{fig:contour_lengthscale}(c). The corresponding equivalent stress is shown in Figs.~\ref{fig:contour_lengthscale}(d)-(f). It is seen that the level of stress in the unit cell increases with increasing length parameter. The critical equivalent stress is seen to increase with increased length parameter in Fig.~\ref{fig:SigC_l3} and the material can therefore withstand higher stresses. 

\begin{figure}[htbp!]
\centering
\begin{overpic}[width=0.5\textwidth]{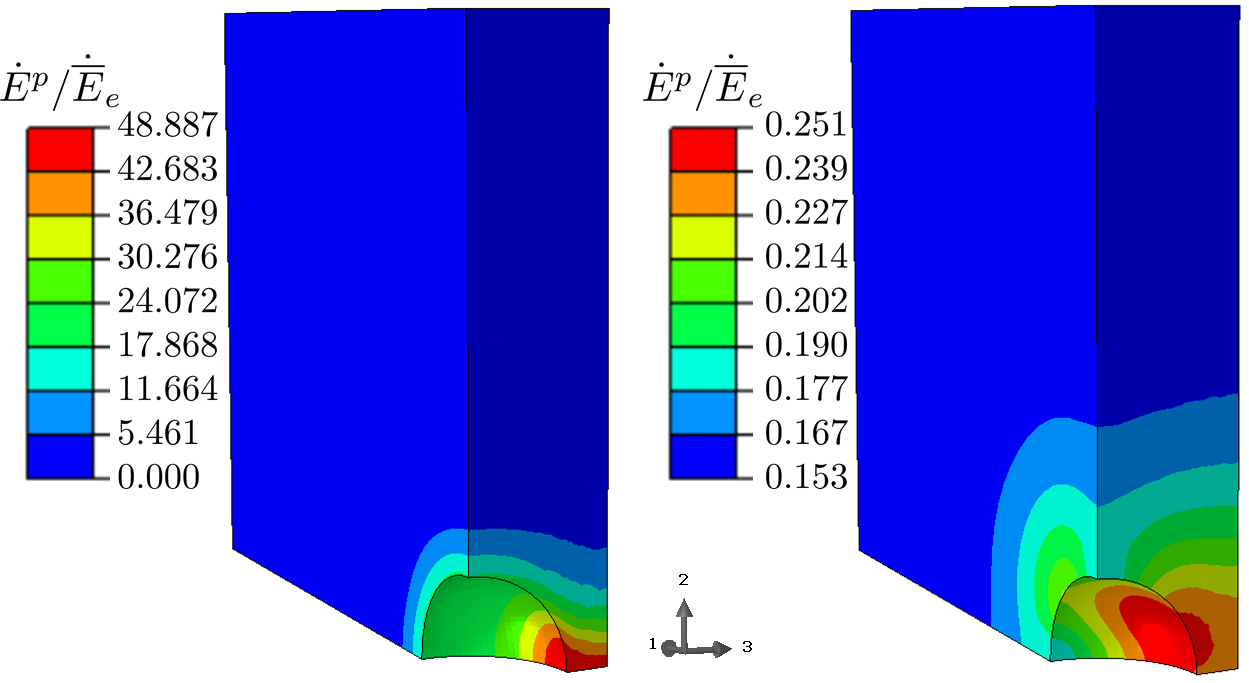}
\put(2,11){(a)} \put(54,11){(b)}
\end{overpic}
\caption{Change in deformation mode with increased length parameter for $l_3/r_0=0.43$ with loading conditions described by $L=1$ and $T=3$. The conventional material with $L_D/r_0=0$ is shown in a), while b) shows a gradient enriched material with $L_D/r_0=1$.}
\label{fig:deformationmode}
\end{figure}

Figure~\ref{fig:deformationmode} shows the change in the deformation mechanism that occurs with increased gradient strengthening. The contour of the normalised rate of equivalent plastic strain, $\dot{E}^p/\dot{\overline{E}}_e$, is shown for the unit cell with the smallest inter-void ligament, $l_3/r_0=0.43$ under loading conditions giving $L=1$ and $T=3$ for the conventional material with $L_D/r_0=0$ and the material with the greatest gradient strengthening contribution, $L_D/r_0=1$. Figure~\ref{fig:deformationmode}(a) shows that plastic deformation has developed and localised in the $l_3$-ligament, as expected for a conventional material at this loading condition. For the matrix surrounding the $l_3$-ligament, plasticity is reduced in favour of localisation in the $l_3$-ligament. However, for the gradient strengthened material, plasticity is not only less developed, in line with the gradient strengthening, but also smeared out across the unit cell, see Fig.~\ref{fig:deformationmode}(b). Localisation is to a little extent observed in the $l_3$-ligament, but overall the entire cell experiences plasticity. This is indicative of a change in deformation mechanism along the lines of the one observed in Ref. \cite{tekouglu2015localization} for a stress triaxiality of 1. However, here it is seen with an increasing length parameter. As $L_D/r_0$ increases, the cell is more likely to undergo simultaneous macroscopic localisation and void coalescence in contrast to a conventional material where the cell predominantly undergoes void coalescence for the same loading conditions.  

The combined effect of the stress triaxiality (for a fixed Lode parameter) and the length parameter is visualised by the rows in Fig.~\ref{fig:SigC_l3}, while the combined effect of the Lode parameter (for a fixed stress triaxiality) and length parameter is visualised by the columns in Fig.~\ref{fig:SigC_l3}. Qualitatively, for a fixed stress triaxiality value, the length parameter has a nearly identical impact for all values of the Lode parameter; the critical stress increases with increasing length parameter. It is, however, interesting that the drop in coalescence stress the cubic unit cell ($l_3/r_0=2.75$) subject to $L=0$ diminishes with increasing length parameter for all values of stress triaxiality, Figs.~\ref{fig:SigC_l3}(d)-(f). This is because increased gradient strengthening delays the intensification of the plastic flow and homogenizes the plastic strain field.

For the lowest stress triaxiality value, $T=1$, the effect of the length parameter is small. Nonetheless, the plastic strain gradients that build up around the void give rise to the small increase in gradient strengthening. For the loading conditions giving $T=1$, the onset of localisation is significantly delayed, thus allowing the material to withstand higher critical stress with a smaller dependence on the inter-void ligament size.  The deformation mechanism prevailing at this low value of triaxiality, where macroscopic localisation and void coalescence occur simultaneously \cite{tekouglu2015localization}, implies that the gradients surrounding the void will not influence the critical equivalent stress to a great extent, as the deformation takes place in the entire unit cell. 

For a higher value of stress triaxiality, $T=2$, the effect of the length parameter is more prominent as seen in Fig.~\ref{fig:SigC_l3}, and the smaller the inter-void ligament size, the greater the effect of the length parameter is. This is because, at higher stress triaxiality values, the plastic flow tends to localise in the inter-void ligaments as the ligaments diminish in size. The localisation induces large plastic strain gradients that in turn contribute to strengthening. The gradient induced strengthening in the inter-void ligament then inhibits further plastic flow localisation and delays void coalescence. Although not shown here, for $L=-1$ and $T=2$, the gradient strengthening is sufficiently large that increasing the value of $L_D/r_0$ from $1$ to $2$, has a negligible effect. For the intermediate length parameter, $L_D/r_0=0.5$, the effects of triaxiality and inter-void ligament size are still visible but greatly reduced due to the smaller degree of gradient strengthening. For the smallest value of the length parameter, $L_D/r_0=0.2$, the critical equivalent stress values follow those of the conventional material, just at a higher relative level for all inter-void ligament size considered. For $L=0$ and $L=1$, for $T=2$, the same effect is seen. The most pronounced effect of the length parameter is seen for $L=-1$, $T=3$ and $l_3/r_0=0.43$ as this configuration has the lowest critical effective stress for the conventional material, but shows the same critical stress for $L_D/r_0=1$ as in the remaining results. 

\section{Summary and conclusions}
\label{sec:Conclusion}
The interaction of the inter-void ligament size and the gradient induced material size effect on void coalescence is investigated for a range of imposed stress states, here characterised by fixed values of the stress triaxiality and the Lode parameter. To this end, three dimensional finite element unit cell calculations for a single initially spherical void embedded in strain gradient enhanced material matrix are carried out. A conventional material matrix (absence of gradient induced strengthening effects) is considered as reference. Increasing the length parameter, and thereby the gradient effect, is equivalent to down-scaling the microstructure. All microstructures considered in this work contain voids that are below a critical flaw size \cite{martinez2019role}. Thus, plasticity theory will reign the material response as the voids are considered too small to be treated as cracks.

The results for the conventional material show that the critical coalescence stress increases when increasing the inter-void ligament size. The effect of the inter-void ligament size is, however, dependent on the imposed stress triaxiality, such that the effect of the inter-void ligament size increases with increasing stress triaxiality. However, above a certain threshold for the inter-void ligament size, the results show a slight decrease in the critical stress. This drop has to do with a transition from plastic flow localisation within the smallest inter-void ligament to plastic flow localization at $\approx 45^\circ$ to the main loading axis. The transition in the plastic flow localisation pattern is found to be particularly pronounced for a Lode parameter of $L=0$. However, irrespective of the Lode parameter value, the transition occurs as the unit cells approach a cubic geometry.

For a void embedded in a strain gradient enhanced material matrix, the value of the critical coalescence stress increases with increasing length parameter i.e. increasing the gradient strengthening effect. The effect of the length parameter is found to intensify with increasing imposed stress triaxiality and decreasing inter-void ligament size. This is due to a propensity for plastic flow localisation in the inter-void ligament when the ligament is small and the stress triaxiality high. Plastic flow localisation introduces large plastic strain gradients which in turn strengthens the ligament and delays further localisation of plastic flow. The strengthening from plastic strain gradients also leads to a weakened dependency in the critical coalescence stress on the inter-void ligament size.  Finally, the results show that there exists a natural upper bound where the gradient strengthening is so severe that the entire matrix material yields. For very large values of the length parameter, the effect of the imposed stress state and the inter-void ligament size vanish, and the critical equivalent stress is identical for all combinations of the unit cell geometry and the loading conditions considered.

%%%%%%%%%%%%%%%%%%%%%%%%%%%%%%%%%%%%%%%%%%%%%%%%%%%%%%%%%%%%%%%%%%%%%%
\begin{acknowledgment}
\noindent This research was financially supported by the Danish Council for Independent Research through the research project ``Advanced Damage Models with InTrinsic Size Effects'' (Grant no: DFF-7017-00121). The computing resources provided by the high performance research computing center at Texas A\&M University are gratefully acknowledged.
\end{acknowledgment}
%%%%%%%%%%%%%%%%%%%%%%%%%%%%%%%%%%%%%%%%%%%%%%%%%%%%%%%%%%%%%%%%%%%%%%
%\clearpage 
%%%%%%%%%%%%%%%%%%%%%%%%%%%%%%%%%%%%%%%%%%%%%%%%%%%%%%%%%%%%%%%%%%%%%%
% The bibliography is stored in an external database file
% in the BibTeX format (file_name.bib).  The bibliography is
% created by the following command and it will appear in this
% position in the document. You may, of course, create your
% own bibliography by using thebibliography environment as in
%
% \begin{thebibliography}{12}
% ...
% \bibitem{itemreference} D. E. Knudsen.
% {\em 1966 World Bnus Almanac.}
% {Permafrost Press, Novosibirsk.}
% ...
% \end{thebibliography}

% Here's where you specify the bibliography style file.
% The full file name for the bibliography style file
% used for an ASME paper is asmems4.bst.
\bibliographystyle{asmems4}

% Here's where you specify the bibliography database file.
% The full file name of the bibliography database for this
% article is asme2e.bib. The name for your database is up
% to you.
\bibliography{asme2e}

%%%%%%%%%%%%%%%%%%%%%%%%%%%%%%%%%%%%%%%%%%%%%%%%%%%%%%%%%%%%%%%%%%%%%%

\end{document}